\documentclass[3p,times,procedia]{elsarticle}
\usepackage{nupha_ecrc}
\usepackage[utf8]{inputenc}
\usepackage{amsmath}

\volume{00}

\firstpage{1}

\journalname{Nuclear Physics A}

\runauth{}

\jid{nupha}

\jnltitlelogo{Nuclear Physics A}

\usepackage{amssymb}
 \usepackage{lineno}
 \usepackage{subcaption}

\usepackage[figuresright]{rotating}

\begin{document}
%\linenumbers

\begin{frontmatter}

%% Title, authors and addresses

%% use the tnoteref command within \title for footnotes;
%% use the tnotetext command for the associated footnote;
%% use the fnref command within \author or \address for footnotes;
%% use the fntext command for the associated footnote;
%% use the corref command within \author for corresponding author footnotes;
%% use the cortext command for the associated footnote;
%% use the ead command for the email address,
%% and the form \ead[url] for the home page:
%%
%% \title{Title\tnoteref{label1}}
%% \tnotetext[label1]{}
%% \author{Name\corref{cor1}\fnref{label2}}
%% \ead{email address}
%% \ead[url]{home page}
%% \fntext[label2]{}
%% \cortext[cor1]{}
%% \address{Address\fnref{label3}}
%% \fntext[label3]{}

\dochead{}
%% Use \dochead if there is an article header, e.g. \dochead{Short communication}
%% \dochead can also be used to include a conference title, if directed by the editors
%% e.g. \dochead{17th International Conference on Dynamical Processes in Excited States of Solids}

\title{Charged-Particle Multiplicity Distributions over a Wide Pseudorapidity Range in Proton-Proton Collisions with ALICE}

%% use optional labels to link authors explicitly to addresses:
%% \author[label1,label2]{<author name>}
%% \address[label1]{<address>}
%% \address[label2]{<address>}

\author{Valentina Zaccolo, on behalf of the ALICE Collaboration}

\address{Niels Bohr Institute, University of Copenhagen, Copenhagen, Denmark}

\begin{abstract}

Multiplicity distributions of charged particles for pp collisions at LHC Run 1 energies, from $\sqrt{s}=$ 0.9 to 8 TeV are measured over a wide pseudorapidity range  ($-3.4<\eta<5.0$) for the first time. The results are obtained using the Forward Multiplicity Detector and the Silicon Pixel Detector within ALICE. The results are compared to Monte Carlo simulations, and to the IP-Glasma model.

\end{abstract}

\begin{keyword}
%% keywords here, in the form: keyword \sep keyword
charged-particle multiplicity distributions  \sep  pp collisions  \sep  saturation models  \sep  forward rapidity

%% MSC codes here, in the form: \MSC code \sep code
%% or \MSC[2008] code \sep code (2000 is the default)

\end{keyword}

\end{frontmatter}

%%
%% Start line numbering here if you want
%%
%%\linenumbers

%% main text
\section{Introduction}
\label{Intro}
The multiplicity distribution of charged particles ($ N_{\text{ch}}$) produced in high energy pp collisions, $\text{P}(N_{\text{ch}})$, is sensitive to the number of collisions between quarks and gluons contained in the colliding protons and, in general, to the mechanisms underlying particle production. In particular, $\text{P}(N_{\text{ch}})$ is a good probe for the saturation density of the gluon distribution in the colliding hadrons. The pp charged-particle multiplicity distributions are measured for five gradually larger pseudorapidity ranges. The full description of the ALICE detector is given in \cite{Aamodt:2008zz}. In this analysis, only three subdetectors are used, namely, the V0 detector \cite{Abbas:2013taa}, the Silicon Pixel Detector (SPD) \cite{Aamodt:2008zz} and the Forward Multiplicity Detector (FMD) \cite{Christensen:2007yc} to achieve the maximum possible pseudorapidity coverage ($-3.4<\eta<5.0$). 

\section{Analysis Procedure}
\label{Analysis}
Three different collision energies (0.9, 7, and 8 TeV) are analyzed here. Pile-up events produce artificially large multiplicities that enhance the tail of the multiplicity distribution, therefore, special care was taken to avoid runs with high pile up. For the measurements presented here, the pile-up probability is of $\sim2\%$. 
%Pile-up contamination happens when 2 or more pp collisions occur in either the same beam crossing or very recent beam crossings. Such events produce artificially large multiplicities that enhance the tail of the multiplicity distribution. For the measurements presented here, the average probability of having more than one interaction in a single bunch crossing where at least one interaction occurs is $\sim2\%$. 
The fast timing of the V0 and SPD are used to select events in which an interaction occurred and events are divided into two trigger classes.
%Analyzed events are triggered by requiring the detection of at least one particle in either the V0A, V0C, or SPD (this is called the MB$_{OR}$ trigger condition). 
The first class includes all inelastic events (INEL) which is the same condition as used to select events where an interaction occurred (this is called the MB$_{\text{OR}}$ trigger condition). The second class of events requires a particle to be detected in both the V0A and the V0C (MB$_{\text{AND}}$ trigger condition). This class is called the Non-Single-Diffractive (NSD) event class, where the majority of Single-Diffractive events are removed. 

The FMD has nearly 100$\%$ azimuthal acceptance, but the SPD has significant dead regions that must be accounted for. On the other hand, interactions in detector material will increase the detected number of charged particles and have to be taken into account. The main ingredients necessary to evaluate the primary multiplicity distributions are the raw (detected) multiplicity distributions and a matrix, which converts the raw distribution to the true primary one. The raw multiplicity distributions are determined by counting the number of clusters in the SPD acceptance, the number of energy loss signals in the FMD \cite{Abbas:2013bpa}, or the average between the two if the acceptance of the SPD and FMD overlap. The response of the detector is determined by the matrix $R_{\text{mt}}$ which, when normalized, is the probability that an event with true multiplicity t and measured multiplicity m occurs. This matrix is obtained using Monte Carlo simulations, in this case the PYTHIA ATLAS-CSC flat tune \cite{d'Enterria:2011kw}, where the generated particles are propagated through the detector simulation code (in this case GEANT \cite{Brun:1994aa}) and then through the same reconstruction steps as the actual data. The response matrix is obtained from an iterative application of Bayes' unfolding \cite{2010arXiv1010.0632D}.
%The finite binning of the measured spectrum will create unphysical fluctuations in the unfolded distribution. To remove them a method based on Bayes’ Theorem \cite{2010arXiv1010.0632D} is used. Bayes' Theorem states that \begin{equation}
%\widetilde{R}_{tm}=\frac{R_{mt}P_{t}}{\sum_{t'}R_{mt'}P_{t'}}\qquad\longrightarrow\qquad U_{t}=\sum_{m}\widetilde{R}_{tm}M_{m}
%\end{equation}
%where $P_{t}$ is an a priori guess of the true distribution and $\widetilde{R}_{tm}$ is the matrix of probabilities that allows one to compute the true multiplicity distribution from the measured one. The unfolded (found) true distribution, $U_{t}$, is then obtained reiterating using $U_{t}$ instead of $P_{t}$ for 10 iterations in this analysis. 
%where $P_{t}$ is an a priori guess of the true distribution and $\widetilde{R}_{tm}$ is the matrix of probabilities that allows one to compute the true multiplicity distribution from the measured one. The unfolded (found) true distribution, $U_{t}$, is then obtained from
%\begin{equation}
%U_{t}=\sum_{m}\widetilde{R}_{tm}M_{m}
%\end{equation}
%the process is reiterated using $U_{t}$ instead of $P_{t}$ for a 10 iterations in this analysis. 
\begin{figure}[tbp]
    \begin{subfigure}[c]{0.39\textwidth}
        \centering
        \includegraphics[width=\textwidth]{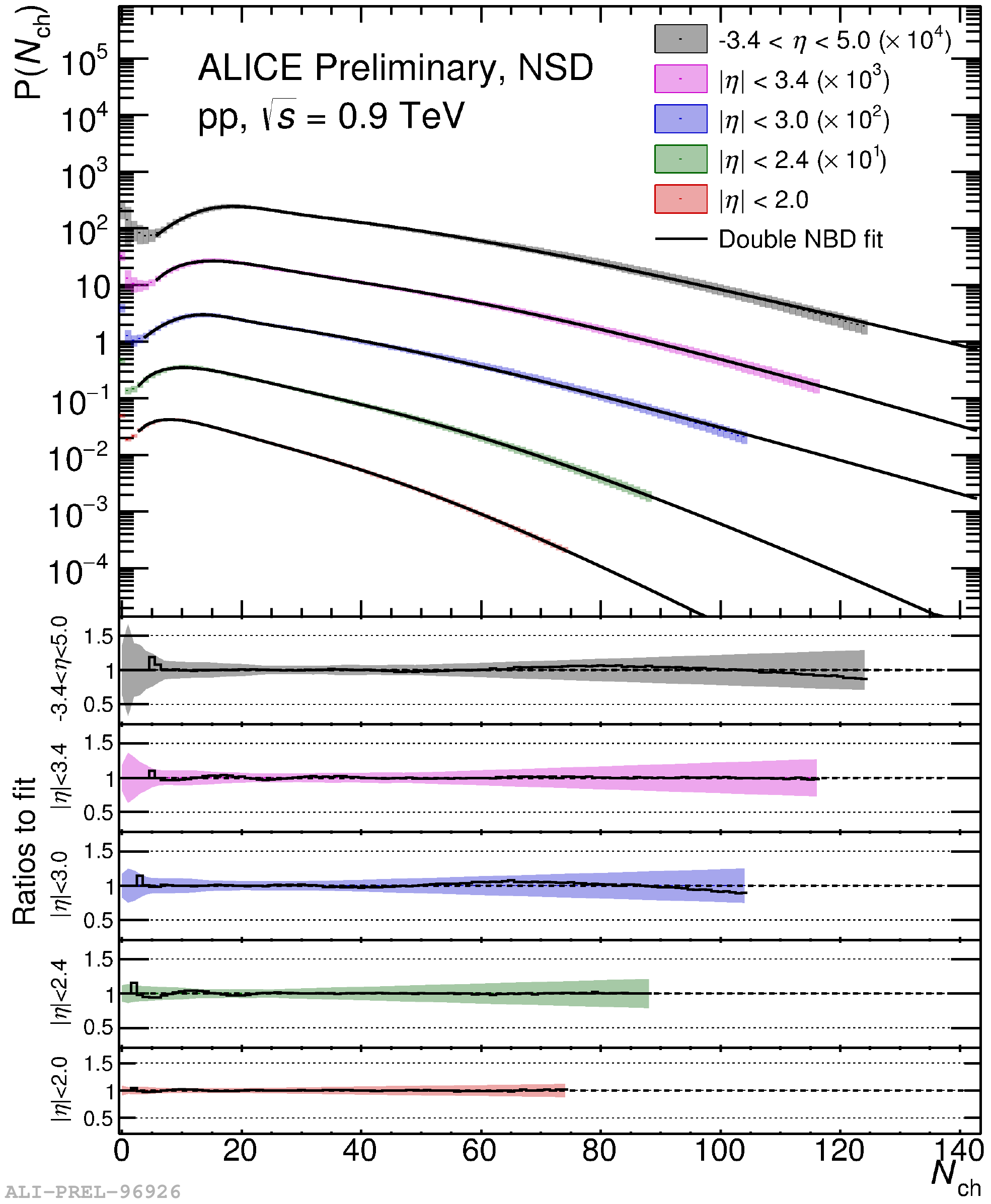}
    \end{subfigure}
\hspace{2cm}    
    \begin{subfigure}[c]{0.39\textwidth}
        \centering
        \includegraphics[width=\textwidth]{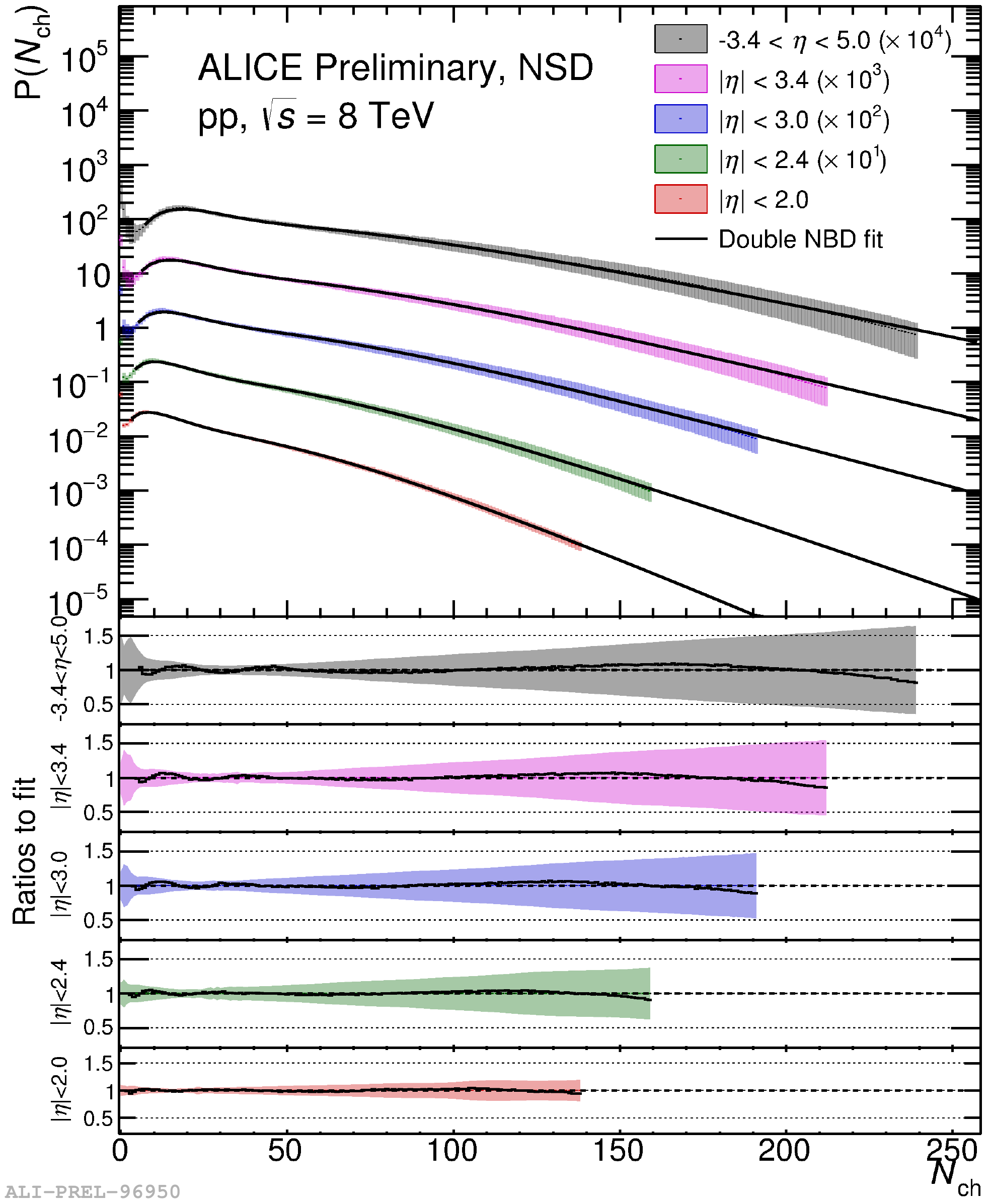}
    \end{subfigure}
\caption{Charged-particle multiplicity distributions for NSD pp collisions at $\sqrt{s}=0.9$ and 8 TeV. The lines show fits to the data using double NBDs (eq.~\ref{eq2}).
Ratios of the data to the fits are also shown.}
\label{V0AND900}
\end{figure}

The probability that an event is triggered, at all, depends on the multiplicity of produced charged particles. At low multiplicities large trigger inefficiencies exist and must be corrected for. The event selection efficiency, $\epsilon_{\text{TRIG}}$, is defined dividing the number of reconstructed events with the selected hardware trigger condition and with the reconstructed vertex less than 4 cm from the nominal IP by the same quantity but for the true interaction classification: $\epsilon_{\text{TRIG}}=N_{\text{ch,reco}}/N_{\text{ch,gen}}$. The unfolded distribution is corrected for the vertex and trigger inefficiency by dividing each multiplicity bin by its $\epsilon_{\text{TRIG}}$ value. Diffraction was implemented using the Kaidalov-Poghosyan model \cite{Kaidalov:2009aw} to tune the cross sections for diffractive processes (the measured diffraction cross--sections at LHC and the shapes of the diffractive masses $M_{\text{X}}$ are implemented in the Monte-Carlo models used for the $\epsilon_{\text{TRIG}}$ computation).

\section{Results}
\label{Results}
\begin{figure}[tbp]
    \begin{subfigure}[c]{0.39\textwidth}
        \centering
        \includegraphics[width=\textwidth]{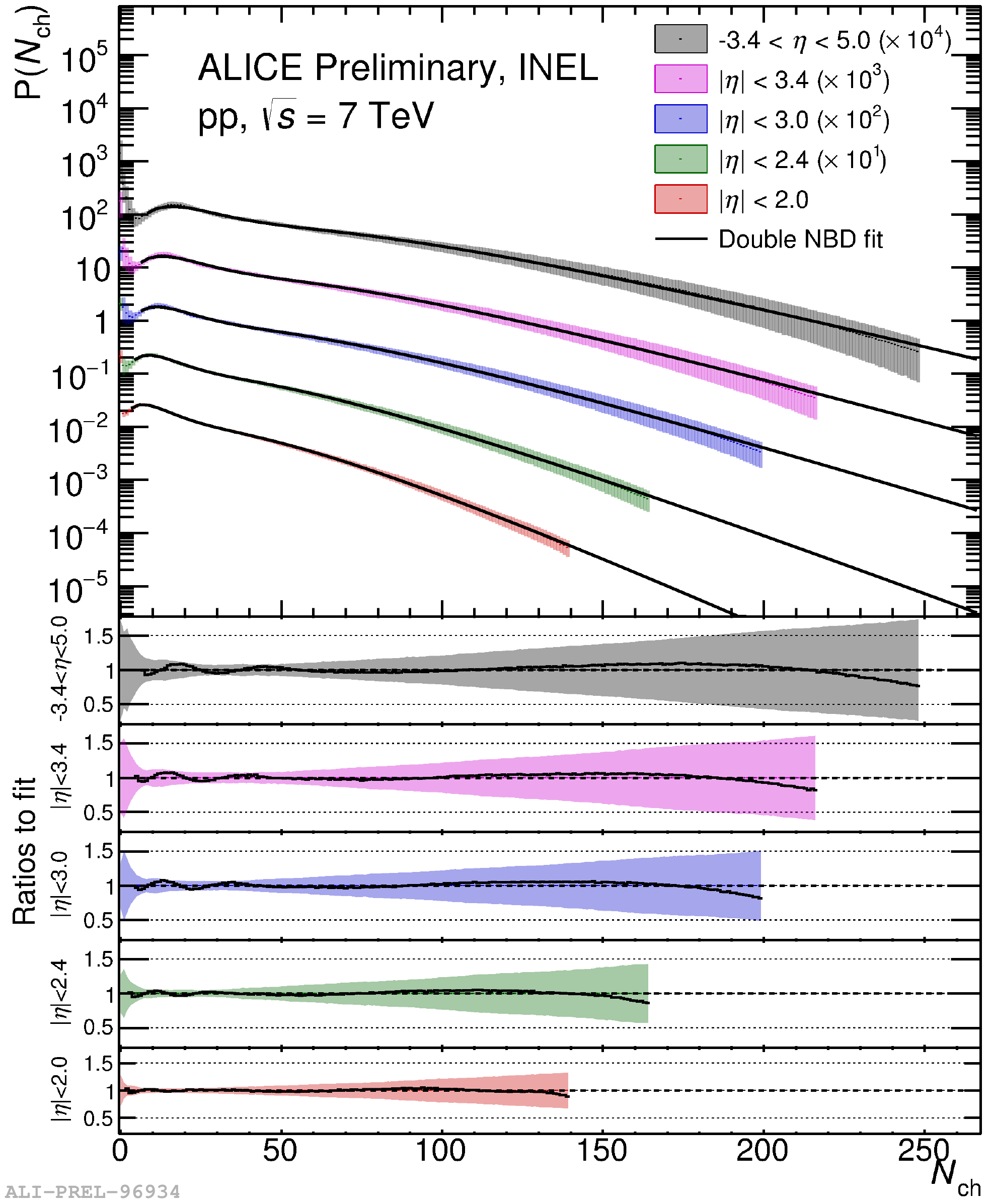}
    \end{subfigure}
    \hspace{2cm}   
    \begin{subfigure}[c]{0.39\textwidth}
        \centering
        \includegraphics[width=\textwidth]{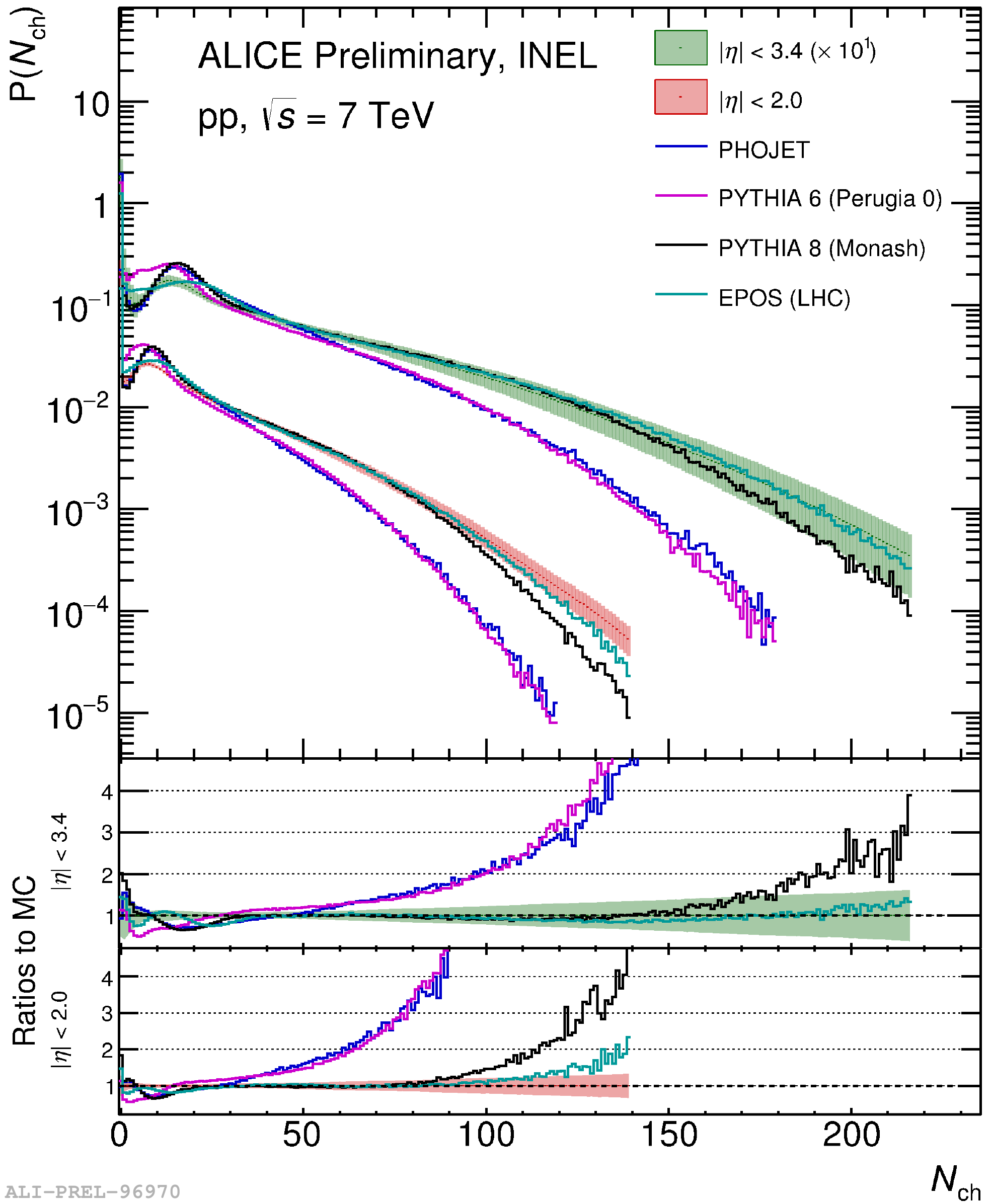}
    \end{subfigure}
\caption{Left: charged-particle multiplicity distributions for INEL pp collisions at $\sqrt{s}=7$ TeV. The lines show fits to the data using double NBDs (eq.~\ref{eq2}). Ratios of the data to the fits are also shown. Right: comparison of multiplicity distributions for INEL events to PYTHIA 6 Perugia 0, PYTHIA 8 Monash, PHOJET and EPOS LHC at 7 TeV.}
\label{V0AND7000}
\end{figure}

The multiplicity distributions have been measured for the two event classes (INEL and NSD) for pp collisions at $\sqrt{s}=$ 0.9, 7, and 8 TeV. Fits to the sum of two Negative Binomial Distributions (NBDs) have been performed here and are plotted together with the results in Figs.~\ref{V0AND900} and \ref{V0AND7000}. The distributions have been fitted using the function
%\begin{align}\label{eq2}
%& P(n)=\lambda[\alpha P_{NBD}(n,\langle n\rangle_{1},k_{1})+(1-\alpha)P_{NBD}(n,\langle n\rangle_{2},k_{2})]  \\ \nonumber
%\text{where}\qquad & P_{NBD}(n,\langle n\rangle,k)=\frac{\Gamma(n+k)}{\Gamma(k)\Gamma(n+1)}\bigg(\frac{\langle n\rangle}{k+\langle n\rangle}\bigg)^{n}\bigg(\frac{k}{k+\langle n\rangle}\bigg)^{k}
%\end{align}
%where
\begin{equation}\label{eq2}
\text{P}(n)=\lambda[\alpha \text{P}_{NBD}(n,\langle n\rangle_{\text{1}},k_{\text{1}})+(1-\alpha)\text{P}_{NBD}(n,\langle n\rangle_{\text{2}},k_{\text{2}})] 
\end{equation}
To account for NBDs not describing the 0--bin and the first bins for the wider rapidities (and therefore removing that bin from the fit), a normalization factor $\lambda$ is introduced. The $\alpha$ parameter reveals the fraction of soft events. It is lower for higher energies and for wider pseudorapidity ranges, where the percentage of semi–hard events included is higher: $\alpha\backsim65\%$ for $\vert\eta\vert<2.0$ at $\sqrt{s}=$ 0.9 TeV and $\alpha\backsim35\%$ for $-3.4<\eta<5.0$ at 7 and 8 TeV. $\langle n\rangle_{\text{1}}$ is the average multiplicity of the soft (first) component, while $\langle n\rangle_{\text{2}}$ is the average for the semi--hard (second) component. The parameters $k_{\text{1,2}}$ represent the shape of the two components of the distribution.

\begin{figure}[tbp]
\centering
\includegraphics[width=0.79\textwidth]{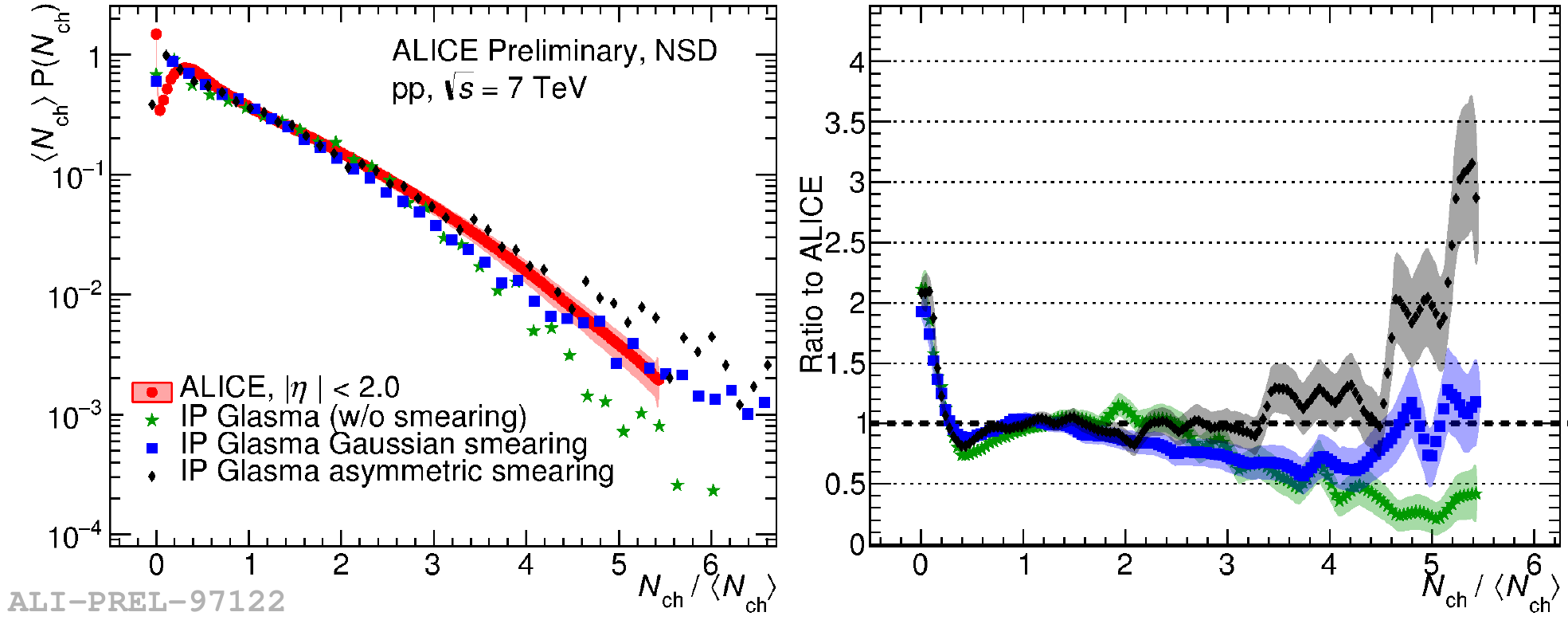}
\caption{Charged-particle multiplicity distributions for pp collisions at $\sqrt{s}=7$ TeV compared to distributions from the IP--Glasma model with the ratio between $Q_{s}$ and the color charge density either fixed (green), allowed to fluctuate with a Gaussian (blue)~\cite{Schenke:2013dpa} or with additional fluctuations of proton saturation scale (black)~\cite{McLerran:2015qxa}.}
\label{IPGlasmaMineNSD7000}
\end{figure}

In Fig.~\ref{V0AND900}, the obtained multiplicity distributions for 0.9 TeV and 8 TeV for the NSD event class are shown for five pseudorapidity ranges, $\vert\eta\vert<2.0$, $\vert\eta\vert<2.4$, $\vert\eta\vert<3.0$, $\vert\eta\vert<3.4$ and $-3.4<\eta<5.0$. The distributions are multiplied by factors of 10 to allow all distributions to fit in the same figure without overlapping.
%\begin{figure}[htbp]
%    \begin{subfigure}[c]{0.45\textwidth}
%        \centering
%        \includegraphics[width=\textwidth]{2015-Sep-21-results_0_1.pdf}
%    \end{subfigure}
%\hspace{1cm}    
%    \begin{subfigure}[c]{0.45\textwidth}
%        \centering
%        \includegraphics[width=\textwidth]{2015-Sep-21-results_2_1.pdf}
%    \end{subfigure}
%\caption{Charged-particle multiplicity distributions for NSD pp collisions at $\sqrt{s}=0.9$ and 8 TeV. The lines show fits to double NBDs.
%Ratios of the data to the fits are also shown.}
%\label{V0AND900}
%\end{figure}
Figure~\ref{V0AND7000} shows the results for the INEL event classes for collisions at 7 TeV (left plot).  Comparisons with distributions obtained with the PYTHIA 6 Perugia 0 tune~\cite{Skands:2010ak}, PYTHIA 8 Monash tune~\cite{Sjostrand:2014zea}, PHOJET~\cite{Bopp:1998rc} and EPOS LHC~\cite{Pierog:2013ria} Monte Carlo generators are shown for INEL events at 7 TeV (right plot) .
Both PHOJET and the PYTHIA 6 strongly underestimate the multiplicity distributions. PYTHIA 8 reproduces well the tails for the wider pseudorapidity range, but shows an enhancement in the peak region. EPOS with the LHC tune models well the distributions, both in the first bins, which are dominated by diffractive events, and in the tails. 
%The evolution of the multiplicity distributions with the center-of-mass energy $\sqrt{s}$ can be studied using the KNO variable $N_{ch}/\langle N_{ch}\rangle$~\cite{Koba:1972ng}. KNO scaling violation is observed as the tails of the resultant distributions increase with increasing energy. Moreover, the violation increases with increasing pseudorapidity range. This behavior was already observed at central rapidities~\cite{Adam:2015gka}. 
The multiplicity distributions are compared to those from the IP--Glasma model \cite{Schenke:2013dpa}. This model is based on the Color Glass Condensate (CGC)~\cite{Iancu:2003xm}. It has been shown that NBDs are generated within the CGC framework~\cite{Gelis:2009wh, McLerran:2008es}. In Fig.~\ref{IPGlasmaMineNSD7000}, the distribution for $\vert\eta\vert<2.0$ is shown together with the IP--Glasma model distributions as a function of the KNO variable $N_{\text{ch}}/\langle N_{\text{ch}}\rangle$. The IP--Glasma distribution shown in green is generated with a fixed ratio between $Q_{s}$ (gluon saturation scale) and density of color charge. This introduces no fluctuations. The blue distribution, instead, is generated with fluctuations of the color charge density around the mean following a Gaussian distribution with width $\sigma=0.09$. The black distributions includes an additional source of fluctuations, dominantly of non-perturbative origin, from stochastic splitting of dipoles that is not accounted for in the conventional frameworks of CGC~\cite{McLerran:2015qxa}. In this model, the evolution of color charges in the rapidity direction still needs to be implemented and, therefore, in the present model the low multiplicity bins are not reproduced for the wide pseudorapidity range presented here.

\section{Conclusions}
\label{Conclusions}
Data from the Silicon Pixel Detector (SPD) and the Forward Multiplicity Detector (FMD) in ALICE were used to access a uniquely wide pseudorapidity coverage at the LHC of more the eight $\eta$ units, from $-3.4<\eta<5.0$.
The charged-particle multiplicity distributions were presented for two event classes, INEL and NSD, and extend the pseudorapidity coverage of the earliest results published by ALICE~\cite{Adam:2015gka} and CMS \cite{Khachatryan:2010nk} around midrapidity, and, consequently, the high-multiplicity reach.
PYTHIA 6 and PHOJET produce distributions which strongly underestimate the fraction of high multiplicity events. PYTHIA 8 underestimates slightly the tails of the distributions, while EPOS reproduces both the low and the high multiplicity events. The Color Glass Condensate based IP--Glasma models produce distributions which underestimate the fraction of high multiplicity events, but introducing fluctuations in
the saturation momentum the high multiplicity events are better explained.

%% The Appendices part is started with the command \appendix;
%% appendix sections are then done as normal sections
%% \appendix

%% \section{}
%% \label{}

%% References
%%
%% Following citation commands can be used in the body text:
%% Usage of \cite is as follows:
%%   \cite{key}         ==>>  [#]
%%   \cite[chap. 2]{key} ==>> [#, chap. 2]
%%

%% References with BibTeX database:

\bibliographystyle{elsarticle-num}
\bibliography{biblio_QM}

%% Authors are advised to use a BibTeX database file for their reference list.
%% The provided style file elsarticle-num.bst formats references in the required Procedia style

%% For references without a BibTeX database:

% \begin{thebibliography}{00}

%% \bibitem must have the following form:
%%   \bibitem{key}...
%%

% \bibitem{}

% \end{thebibliography}

\end{document}